\documentstyle[aps,prb,epsf]{revtex}
\begin{document}

\twocolumn[\hsize\textwidth\columnwidth\hsize\csname
@twocolumnfalse\endcsname

\title{Core-softened potentials and the anomalous properties of water}
\author{E. A. Jagla}
\address{The Abdus Salam International Centre for Theoretical
Physics (ICTP), I-34014 Trieste, Italy}
\maketitle
 
\begin{abstract}
We study the phase diagram 
of a system of spherical particles interacting
in three dimensions through a potential consisting of a strict hard
core plus a linear repulsive shoulder at larger distances.
The phase diagram (obtained numerically, and analytically in a
limiting case)
shows anomalous properties that are similar to those observed in
water. Specifically, we find maxima of density and isothermal
compressibility as a
function of temperature, melting with volume contraction, and
multiple stable crystalline structures. If in addition a long
range attraction between the particles is included,
the usual liquid-gas coexistence curve with its critical point is obtained.
But more interestingly, 
a first order
line in the metastable fluid branch of the phase diagram appears, ending
in a new critical point, as it was suggested to occur 
in water. In this way the model provides a comprehensive, 
consistent and unified picture of most of the anomalous thermodynamical
properties of
water, showing that all of
them can be qualitatively explained by the existence of two competing
equilibrium values for the interparticle distance.
\end{abstract}
 
\vskip2pc] \narrowtext

\section{introduction}

Water is an anomalous substance in many respects.\cite{agua} Liquid water
has a
maximum as a function of temperature in both  density and isothermal
compressibility. It solidifies with volume increasing at low
pressures, and the 
solid phase (ice) shows a remarkable variety of crystalline
structures in different sectors of the
pressure-temperature plane. Some of these properties are known
from long 
ago, but their origin is still controversial. In an effort to rationalize
these anomalous properties, the supercooled (metastable) sector of the
phase diagram
of liquid water has received much attention in the last
years.\cite{angell82,stanleyrev,deb,angell0,speedy0} It was
observed that when appropriately cooled (using
techniques for preventing crystallization) water becomes a viscous 
fluid with many properties (as heat capacity and isothermal
compressibility) displaying a tendency that has suggested even a
thermodynamic singularity at some lower temperature.\cite{spang} 
Although there is a limit of about 235 K below which water cannot be
cooled without crystallization, amorphous states of water at much lower
temperatures can be obtained by different techniques. All these amorphous
states are observed to correspond to one of two different structures
(referred to as low-density amorphous -LDA- and high-density amorphous -HDA)
that differ by 
about 20 per cent in density, which transform reversibly one into the
other upon changes of pressure.\cite{mishima1} There is evidence that
these amorphous states are thermodynamically connected with fluid water,
although a direct verification is not possible due to recrystallization at
intermediate temperatures.\cite{scott99}

The observation of LDA and HDA was an experimental clue that 
led to the proposal of the second critical point
hypothesis.\cite{stanleyrev,stanley3,stanley2} This hypothesis states that
in the
deeply supercooled region water can exist in two different amorphous
configurations, separated by a line of first order transitions.
This line should end in a critical point very much as the usual
liquid-vapor line ends
in a critical point. This hypothesis, in addition to obviously explain the
reversible transformation between LDA and HDA, provides a natural though
phenomenological explanation for the
anomalous behavior of density and isothermal compressibility. However, the
very existence of the second critical point is known to be not
neccesary for the appearance of other anomalies,\cite{tanaka3,stanley4}
and the issue 
of what are the microscopic properties of water molecules that may
produce the appearance of the second critical point are only poorly
understood. In all cases it seems to be
crucial the fact that water (because of the particular form of its   
molecules and peculiarities of the hydrogen bond) exhibits competition between
more expanded structures (preferred at low pressures) and more compact
ones (which are favored       
at high pressures). But it is not obvious to what extent this
simple fact can be made responsible for all the anomalies of water,
or if more subtle properties of the interaction potential (in
particular, cooperative hydrogen bonding)\cite{stanley4,texeira} are
crucial.

Numerical simulations based on some of the available pair potentials for
the interaction between water molecules reproduce reasonably well many of
its properties, although the systems that have been studied are
strongly limited in size, due to computational
constraints.\cite{stanleypc,tanaka} These simulations only suggest the    
existence of the second critical point, but up to now they were not able  
to prove its existence unambiguously.
Other simplified
and in some cases ad-hoc models have been devised to show the appearance of
anomalous properties in the phase diagram.\cite{otros} Some of these
models have a second critical point, but in these cases a global
characterization of the phase diagram that includes all other anomalies
has not bee achieved.
In all cases, the models used have as a
fundamental ingredient the competition between expanded, less dense
structures and compressed, more dense ones. 

It is the goal of the present work to show that a very simple model of
spherical particles
interacting trough a repulsive potential that possesses two different
preferred equilibrium positions has a phase
diagram in which a) lines of maxima for the density and the isothermal
compressibility of the liquid exist; b) the fluid phase freezes
with an
increase in volume in some pressure range; c) the solid phase has multiple
different crystalline structures depending on $P$ and $T$. When a long range
van der Waals attraction is included on top of the previous, exclusively
repulsive potential, the system d) preserves the anomalies
existent in the non-attractive case; e) develops a liquid gas first order
coexisting line that ends in a critical point in the usual fashion; f)
depending on the strength of the attractive potential, a line of first order
transitions separating two amorphous phases in the supercooled region
appears.
This line ends in a second critical point from which the line of maxima
in isothermal compressibility starts.

These statements will be justified mostly using numerical (Monte Carlo)
techniques. However, for the deep supercooled states, the long
equilibration times make the numerical studies not completely reliable. In
this case,
the numerical results are supported by analytical calculations in a limiting
case of the interacting potential that shows neatly how the second critical
point appears.

The paper is organized as follows. In Section II we give a brief
description of the interaction potential and the numerical technique.
Section III focus on the different stable crystalline configurations of
the system. In Section IV we study the phase diagram of the fluid phase,
both where it is thermodynamically stable and also in the supercooled
region. Here we rely both in analytical calculations and in simulations,
and show how a second critical point can appear. In Section V we describe
the melting of the most expanded solid structure and its anomalies.
Finally, in Section VI we take all the results as a whole and comment upon
their importance for the understanding of the properties of water.

\section{Model and numerical details}

The interaction potential $U(r)$ between particles that we will consider 
is chosen to be the hard-core plus linear-ramp potential originally
studied by Stell and Hemmer.\cite{stell1}
The radius of the hard
core is taken to be $r_0$, and the ramp extends linearly from
the value $\varepsilon_0$ at $r=r_0$ up to 0 at $r=r_1$. In addition, a long
range van der Waals attraction will be included through a global term in
the energy
per particle of the system proportional to $-\gamma/v$, with $v$ the
specific
volume and $\gamma$ a coefficient that represents the total integrated
strength of the attraction.
The van der Waals term can be accounted for
without its explicit inclusion in the simulations in the following
way.\cite{hemmerleb}
Since the free energy per particle contains the term $Pv-\gamma/v$,
when minimizing with respect to $v$ the combination $P+\gamma/v^2$
appears. So we will call $P^*\equiv P+\gamma/v^2$, and make the
simulations in terms of $P^*$, with $\gamma=0$. At the end, the self
consistent replacement $P^*\rightarrow  P+\gamma/v^2$ is made, and this
provides the results for finite $\gamma$.

Numerical simulations are performed at constant $P$, $T$, and $N$
(the number of particles) by standard Monte Carlo techniques. Periodic boundary
conditions are used in the three directions. The equilibrium volume at each
pressure is reached by allowing the system size to increase or decrease
through Monte Carlo movements that expand of contract all coordinates of the 
particles as well as the total size of the system. The rescaling is
accepted of rejected depending on the energy change it involves. The
contraction-expansion procedure is made independently for the three
spatial coordinates, and the 
maximum ratio between the size of the system in different directions is
limited to 1.2. 

A subtle but important technical change was introduced in the simulation
procedure to be able
to equilibrate the volume in the low temperature region. Let us think for
instance
of the $T=0$ case. If we increase $P$, as soon as two particles are at a
distance $r_0$ from each other, the total volume gets stuck in the simulation
because (due to the scheme adopted for doing volume changes) the volume
can be reduced further only if a global contraction of all
coordinates reduces the energy. But this contraction would bring the two
particles (which are already at a distance $r_0$) at a distance lower
than $r_0$, then formally to a state of infinite
energy, and so the trial movement is rejected. To avoid this problem we
relax a
bit the rigid hard core at $r_0$, replacing it by a new linear ramp between
$r=0$ and $r=r_0$ of the form $U(r)=\varepsilon_0 [50(1-r/r_0)+1]$. The
last term
is included in order to match smoothly the potential for $r>r_0$. This
modification was seen not to modify the behavior of the system,
it just provides a convenient way of reaching the equilibrium values of 
the volume within a reasonable computing time in the low temperature
regime.

The property of the potential that renders it interesting
for our problem is that depending on the external force acting on them,
two particles
prefer to be at distance $r_0$ or $r_1$ from each other, and the effect of
this simple fact on the phase diagram is dramatic.\cite{sadr}
Already in the original papers about this potential,\cite{stell1} it was
realized that in
fact the competition between configurations with particles at
distances
$r_0$ and $r_1$ may produce the appearance of polimorfism in the system. More
precisely, in 1D the system may exhibit many liquid phases when
$\gamma\neq 0$, with sharp transitions between them. In 3D
these transitions occur within the solid region of the phase diagram, and it
was suggested that they may still be observable as isostructural transitions
of the solid.\cite{stell1,stell2} In this connection we want to emphasize
the
following two points:
i) isostructural transitions within the solid phase for this kind of
potentials  are usually preempted by the appearance of new intervening
solid phases of
different symmetry; ii) polimorfism of the liquid
state as observed in the 1D system also appears in 3D samples, but now
in the supercooled liquid state, and is the responsible for the existence of
the second critical point. We address these two points in the next two
sections.

\section{Crystalline configurations at $T=0$}

Multiple crystalline structures for our potential
arise from the competition between expanded and contracted structures. 
At $T=0$ the preferred configuration of the system will be the one that
minimizes the enthalpy $h\equiv e +Pv$. At low $P$ the best
way of minimizing $h$ is first to minimize $e$, and then $v$. This is
achieved in a close
packed structure with first neighbors distance equal to $r_1$. At very
large $P$, in order to minimize $h$ it is energetically more convenient
to minimize $v$, and the structure becomes again a close packed one with
first neighbors distance equal to $r_0$. However, in  the range in which
these two configurations have approximately the same enthalpy, there
are others which are more stable, having pairs of neighbor particles both
at distances $r_0$ and $r_1$. For our potential $U(r)$ (and also for more
general  potentials) the existence of other stable configurations 
can be demonstrated quite generally.\cite{jaglapre,jaglajcp} However, to
tell safely which is
the structure of lowest enthalpy for each $P$ is a problem for
which a
closed solution is not known. The usual approach is to compare the
enthalpies of
structures proposed beforehand, using simulated
annealing techniques to guess the possible structures.
We refer to [22,23] for a detailed discussion of the
two-dimensional case,
and also for
discussions on the neccesary conditions on the potential for different
structures to appear. 

\begin{figure}
\narrowtext
\epsfxsize=3.3truein
\vbox{\hskip 0.05truein
\epsffile{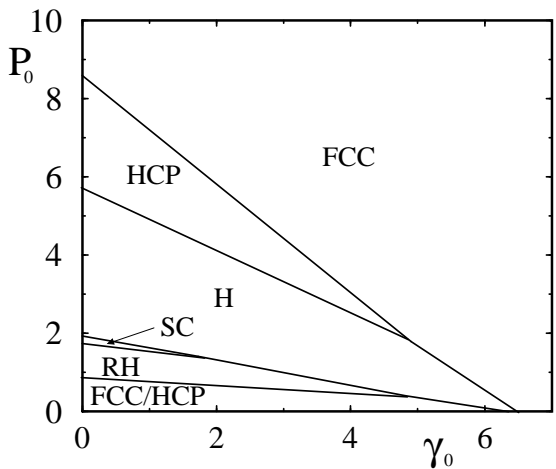}}
\medskip
\caption{Crystalline structures of highest stability for different values 
of $P$
and $\gamma$, for $r_1/r_0=1.75$ and $T=0$ (adimensional values of $P$
and $\gamma$ are defined as $P_0\equiv Pr_0^3 \varepsilon_0^{-1}$, and
$\gamma_0\equiv \gamma\varepsilon_0^{-1} r_0^{-3}$. The search was
performed
among structures of
the cubic (simple (labeled SC), face centered (FCC), and body centered),
tetragonal
(simple, and body centered), rhombohedral (RH), and hexagonal (simple
(H), and close packed (HCP)) crystalline systems, with only one
particle per unit cell (except for the HCP structure). However, other more
complex structures
cannot be ruled out.}
\label{gamap}
\end{figure}

Here we show only in Figure \ref{gamap} the result
of comparing (for $r_1/r_0=1.75$) the enthalpies of particles arranged in 
Bravais lattices
corresponding to cubic, tetragonal, rhombohedral, and hexagonal
systems as a function of
$P$ and $\gamma$. The structures searched are those that can be defined by
no more than two parameters (that fix the form and size of the Bravais
lattice). All of them were supposed to have only
one particle per unit cell, except for HCP (which has two), that was
included due to its known stability.
We find 
five different crystalline configurations as a function
of $P$. Note also how the increasing of the van der Waals attraction moves
all the
borders between structures to lower pressures, since those
that are stable at higher $P$ are always more
compact, and thus become more stable in the presence of the
van der Waals term.
It must be kept in mind that there may be other
configurations (corresponding to other crystalline systems, or with more
complex unit cells) with lower enthalpy. In fact, they are likely   
to occur, as for instance in the 2D case crystalline structures with up to
five particles per unit cell appear. \cite{jaglapre,jaglajcp} Only a
thorough numerical work can determine all possible structures.

\section{Properties in the Fluid and Supercooled Regions}

In the previous section we saw that at $T=0$ there is a sequence of solid
phases interpolating between the lowest density and highest density
ones. At each transition (as $P$ is increased) a finite fraction of
particles that were at distance $r_1$ from each other passes to be at
distance $r_0$. Each of these rearrangements involves a change of symmetry,
and thus
the appearance of a new crystalline structure. The picture is different in
the metastable, disordered sheet of the phase diagram. 
At very low pressures the
particles behave as hard spheres
with radius $r_1/2$. Hard
spheres are known to have a maximum density of
random packing, corresponding to a volume per particle $v_1$ than in 
our case is $v_1\simeq 0.808 r_1^3$. Being
the densest disordered structure
possible for hard spheres, this is also the thermodynamically stable
amorphous configuration of the system when $T=0$, namely, the one which
minimizes the enthalpy. When $P\rightarrow
\infty$ the linear ramp of $U(r)$ is irrelevant to calculate the free
energy, and the
thermodynamically
stable configuration is again a random packing of spheres, now with radius
$r_0/2$. As in the crystalline case, the nearest neighbor distance between
particles must collapse from 
$r_1$ to $r_0$ as a function of $P$.
The crucial question is whether this collapse is 
discontinuous  at some well defined $P$ (or even if there are more
than one transitions at different values of $P$) or if it is just a smooth
crossover. We address the issue
in the following two subsections, first analytically (when $r_1/r_0
\rightarrow \infty$) and then numerically, but 
let us quote briefly the answer to this question in advance. For the case
of no
van der Walls attraction the behavior of the specific volume $v$ as a function
of $P$ is smooth at any finite temperature. But there is a range of pressures
in which $\frac{\partial v}{\partial P}$ is anomalously large. This fact is
enough for the van der Waals interaction to produce (if sufficiently
strong) 
the appearance of a metastable critical point and a
line of first order transitions between two disordered
structures with a finite difference in density.

\subsection{The limit $r_1/r_0\rightarrow \infty$}

We will start by considering only the  
repulsive part of the potential (i.e., $\gamma=0$).
As we already said, at very low pressures the particles behave as hard
spheres with radius $r_1/2$. At $T=0$ the enthalpy per particle of this
configuration is
$h=Pv_1$. This is the thermodynamically stable state upon
increasing $P$ up to the point where it is
energetically more convenient to overlap neighbor particles in pairs. The
structure will now be similar to the low pressure one, but with two
particles overlapped on each position (see a sketch of this fact in Figure
\ref{t0}).
The enthalpy of this configuration is
$h'=Pv_1/2+\varepsilon_0/2$, since now the total volume of the system is
reduced in a factor 2, and an energy $\varepsilon_0$ must be counted for
each
pair of particles. The pressure at which $h=h'$ determines the
transition pressure $P_{TR}=\varepsilon_0/v_1$. Close to the point
($P_{TR}$, $T=0$) of the
phase diagram, we can calculate approximately the free energy of the system
in the
following way. Let us suppose we have $N$ particles, $n$ of them in
non overlapped positions  and $n'$ pairs of overlapped particles
($N=n+2n'$). The configurational free energy of the system
may be written as (higher than double overlaps that will ocurr at higher
pressures are dismissed)
 
\begin{equation}
F=[Pv'-Ts^{HS}(v')](n+n')+\varepsilon _0 n' -Ts^I
\end{equation}
where $v'\equiv V/(n+n')$ [note that $v'$ is not the specific volume, which
instead is given by $v=V/(n+2n')$],
$s^{HS}(v')$ is the entropy per particle of hard spheres with radius
$r_1$ on the metastable sheet, and $s^I$ is the configurational entropy
for choosing which particles will be in pairs, and which ones will be
singled
$\left [s^I=k_B\ln \left ( 
\begin{array}{c}
n+n' \\ n'
\end{array}
\right ) \right]$. Using $v'$ and $n'$ as
independent variables for minimizing $F$ we obtain the
equations
\begin{eqnarray}
\frac{P}{T}&=&\frac{\partial s^{HS}(\tilde v')}{\partial \tilde v'}\label
{algo0}\\
\frac{P\tilde v'}{T}-s^{HS}(\tilde v')&=&\frac{\varepsilon_0}{T}-k_B \ln %
\frac{\left ( 1-2n'/N \right )^2 }{\left ( 1-n'/N \right )n'/N}
\label{algo}
\end{eqnarray}
(we use $\tilde v'$ in this case with $\gamma=0$, to distinguish from the
$\gamma \neq 0$ case).
 
The first one is the equation of state of
hard spheres in the metastable region. We will
use for it the following expression provided by Speedy\cite{speedy}
\begin{equation}
\frac{P}{T}=\frac{2.65k_B}{\tilde v'-v_1}
\label{sp}
\end{equation}

For given values of $P$ and $T$, $\tilde v'$ is determined from this
equation, and the value obtained is used to determine $n'$ from
(\ref{algo}).\cite{nota}

\begin{figure}
\narrowtext
\epsfxsize=3.3truein
\vbox{\hskip 0.05truein
\epsffile{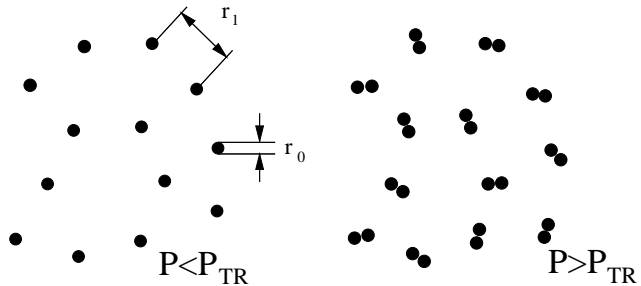}}
\medskip
\caption{Sketch of two dimensional configurations of particles in a
glassy state at $T=0$ for
$r_1/r_0$ large, although finite, below and above the transition pressure
$P_{PR}$. Drawings were made so as to emphasize that for $P>P_{TR}$ (b)
the 
structure is similar to that at $P<P_{TR}$ (a) but with two particles per
site, instead of one (this fact is strictly valid if
$r_1/r_0\rightarrow
\infty$). The density of the system is roughly twice in (b)
than in (a). The energy per particle is 0 in (a) and $\varepsilon_0/2$ in
(b).}
\label{t0}
\end{figure}
  
The volume per particle of the system is $\tilde v\equiv V/N= \tilde
v'(n+n')/N$.
Although $\tilde v'$
has a behavior on $P$ and $T$ that is the same as for hard spheres, the
$(n+n')$ factor (that takes the value $N$ when $T=0$, $P<P_{TR}$, and $N/2$
when $T=0$, $P>P_{TR}$) makes the behavior of $v$ on $T$ and $P$ be
non
trivial. We see the surface  $\tilde v(P,T)$ that is obtained from
equations
(\ref{algo0}), (\ref{algo}), and (\ref{sp}) in Fig. \ref{vpt1}. 
At $T=0$ we obtain the expected result, with $\tilde v(P,T=0)$ passing
from
$v_1$ to
$v_1/2$ at the transition pressure $P_{TR}$.
But at any finite $T$ the entropy transforms this jump in a smooth
crossover.
The point $P=P_{TR}$, $T=0$ is for this system the metastable
critical point.

The inclusion of a finite long
range attraction through a non zero $\gamma$ produces the critical
point to move into the $T>0$ region. The mechanism is identical to the one
that produces the appearance of the usual liquid-gas coexistence curve.
We must replace $P$ by $P+\gamma/v^2$ in
expressions (\ref{algo0}), (\ref{algo}), and (\ref{sp}) to take account of
the van der Waals attraction.
A singularity in $v(P)$ at a finite temperature exists if
$\partial v/\partial P$ becomes negative (this is the signature of a van
der
Waals loop, and thus of a first order transition). Since
\begin{equation}
v(P)=\tilde
v(P+\gamma/v^2),
\label{vvtilde}
\end{equation}
we can calculate $\partial v/\partial P$ as
\begin{equation}
\partial v/\partial P=\left .\frac{\partial \tilde v(x)/\partial x} {1+
\frac{2\gamma}{v^3}
\frac{\partial \tilde v(x)}{\partial x}}\right|_{x=P+\gamma/v^2}.
\label{pc}
\end{equation}
We see that a singularity occurs if  $\partial \tilde v(P)/\partial P$
is larger (in absolute
value) than $\frac{v^3}{2\gamma}$. This always happens in our model close
enough to $P_{TR}$, $T=0$.
In Figure \ref{vpt2} the function $v(P,T)$, calculated using
the self consistency condition (\ref{vvtilde}), with $\gamma_1=0.1$
($\gamma_1\equiv \gamma\varepsilon_0^{-1}r_1^{-3}$)is
shown.  

\begin{figure}
\narrowtext
\epsfxsize=3.3truein
\vbox{\hskip 0.05truein
\epsffile{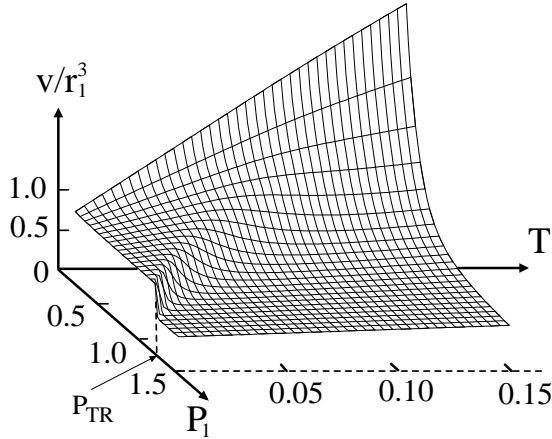}}
\medskip
\caption{The surface $v(P,T)$ for $\gamma_1=0$ ($\gamma_1\equiv
\gamma\varepsilon_0^{-1}r_1^{-3}$), when
$r_1/r_0\rightarrow\infty$. Only the $T=0$ isotherm is singular, at
$P=P_{TR}$ ($T$ is measured in units of $k_B^{-1}\varepsilon_0$, 
$P_1\equiv Pr_1^3\varepsilon_0^{-1}$).}
\label{vpt1}
\end{figure}

\begin{figure}
\narrowtext
\epsfxsize=3.3truein
\vbox{\hskip 0.05truein
\epsffile{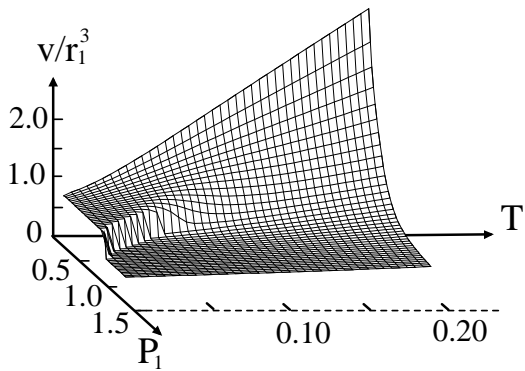}}
\medskip
\caption{Same as Figure (\ref{vpt1}), but with $\gamma_1=0.1$. Now a
discontinuity exists for all $T{\protect\lesssim} 0.37$.}
\label{vpt2}
\end{figure}  

The rapid change in $v$ as a function of $P$ close to the critical point
is the responsible for the anomalous behavior of $v$ and the isothermal
compressibility $K_T\equiv -\frac{1}{v}\frac{\partial v}{\partial P}$. We
see the
location in the $P$-$T$ diagram of the extrema of $v$ and $K_T$ as a
function of temperature in Figure \ref{pt1}. We also see in this figure
the first order line that appears due to the van der Waals attraction,
ending in the critical point C', and also the two spinodal lines that mark
the limit of metastability of the two phases on both sides of the
first order line. Note that the singularity of the thermodynamic
properties
at the critical point manifests itself in anomalous properties (of $v$ 
and $K_T$) that can be
detected at higher temperatures.

\begin{figure}
\narrowtext
\epsfxsize=3.3truein
\vbox{\hskip 0.05truein
\epsffile{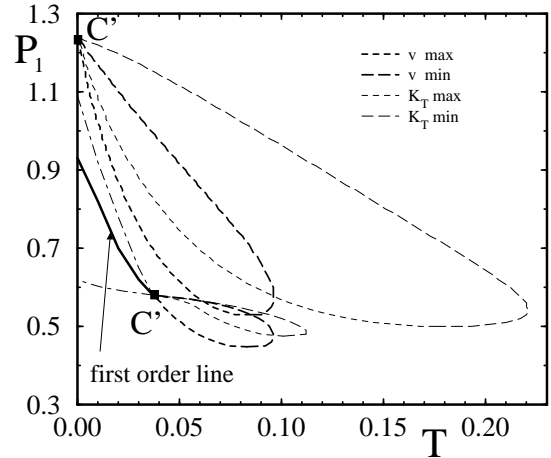}}
\medskip
\caption{The locus of extrema of $v$ and $K_T$, calculated from the
$v(P,T)$
function, for $\gamma_1=0$ (upper curves) and $\gamma_1=0.1$ (lower curves).
For finite $\gamma_1$ a
first order line ending in a critical point $C'$ appears. The
dashed-dotted lines are the spinodals of this first order transition.}
\label{pt1} 
\end{figure}

The analytical treatment of the
case $r_1/r_0 \rightarrow \infty$ provides insight into the appearance of
the second critical point in the phase diagram of water. In fact, the
anomalies in $v$ and $K_T$ exist
even for exclusively repulsive potentials. It is the van der Walls
attraction that brings the critical point to a finite temperature, in the
same way that it is this attraction (or a more realistic finite range one)
that generates the familiar liquid-gas coexistence line. Now we
will see
how much of this scenario remains for finite $r_1/r_0$.

\subsection{Numerical results for finite $r_1/r_0$}

When $r_1/r_0$ is finite, no analytical calculation seems to be possible to
tell the existence or not of the metastable critical point. But
guided by the previous findings, we can more safely interpret the
numerical results.

In Figure \ref{vt1} we show the results of numerical simulations for
$r_1/r_0=1.5$.
Rapid runs (2000 steps per temperature)
were made by decreasing the temperature at different values of $P$
in a system of 197 particles.
The rapid cooling allows to reach the supercooled states without
crystallization except in the continuous-dashed region.\cite{tanaka2} 
The curves shown are averages over 20 different runs. Only
the points in which the system did not crystallize and displays well
reproducible values for the density are shown. In spite of this, since the
runs were rapid and the low temperature states are highly viscous, we can
rise some doubts about the final state reached for
$T\rightarrow 0$. It might be that we are observing some frozen
configuration typical of larger $T$. To answer this point
we made runs at a low temperature ($T=0.01$) increasing and
decreasing pressure (Fig. \ref{vdept0}). The results of this
simulation show evident effects
of hysteresis due to the glassiness of the states. This hysteresis was
seen not to be greatly reduced by decreasing the rate of temperature
change in a factor ten. But anyway the hysteresis path
encloses the values of $v$ obtained by decreasing $T$ at fixed $P$ (large 
symbols in Fig. \ref{vdept0}), and we have also checked that the radial distribution
functions are comparable in both cases.
The finding of essentially the same results when we
arrive from different paths in the
$P$-$T$ plane is an indication that in fact these are 
thermodynamic values. 

\begin{figure}
\narrowtext
\epsfxsize=3.3truein
\vbox{\hskip 0.05truein
\epsffile{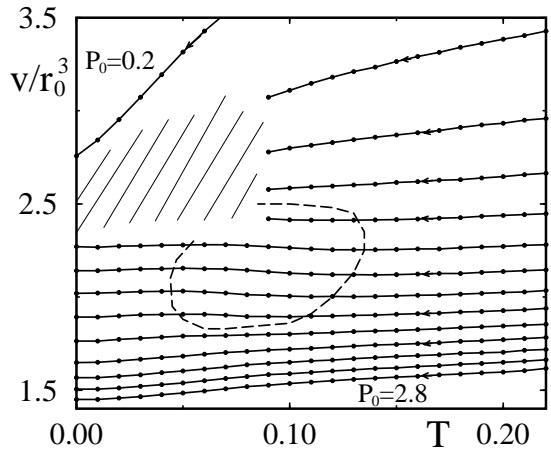}}
\medskip
\caption{Specific volume as a function of $T$ for different values of $P_0$
($P_0=Pr_0^3\varepsilon_0^{-1}$) from 0.2 to 2.8 in steps of 0.2,
from the numerical
simulations with $r_1/r_0=1.5$. Simulations were done reducing rapidly the
temperature from the fluid state. Averages over 20 runs are shown.
The continuous-dashed region corresponds to points where the system
crystallization could not be avoided even in these rapid runs,      
and are thus not included. Within the region limited by the dashed line
$\partial v/\partial T$ is negative.}
\label{vt1}   
\end{figure}

There is
no sign in Fig. \ref{vt1} (in contrast to the $r_1/r_0 \rightarrow \infty$
case) of an abrupt jump in $v$ as a function of $P$ at $T=0$, all that
remains is a value of
pressure with a maximum in
$\partial v(P,T=0)/ \partial P$ (as is seen in Fig. \ref{vdept0}, close to
$P_0=2$). Around this value $K_T$ has a maximum (as a function of $T$) at
$T=0$, whereas $v$ has maxima
and minima at finite temperature, as can be seen from Fig. \ref{vt1} in the range
$1.0\lesssim P_0\lesssim 1.8$ (on the dashed
line). These
facts are sufficient for the van der Walls
attraction to induce the appearance of a critical point, if $\gamma$ is large
enough.
In fact, we find that for $\gamma_0\equiv\gamma_0^{cr}\simeq 4.2$ a critical point
enters the phase diagram at
$T=0$, $P_0\simeq 0.65$. The location of the critical point as a function of
$\gamma$ can be determined from data as those of Fig. \ref{vt1} by requiring that
$\partial v/\partial P\rightarrow \infty$,  and $\partial^2 v/\partial
P^2\rightarrow \infty$ calculated according to Eq. (\ref{pc}).
We find in our case that for $\gamma_0$ slightly larger than $\gamma_0^{cr}$ the
critical point position can be estimated as

\begin{eqnarray}
T^{cr}&\simeq& 0.07(\gamma_0-4.2)\\
P_0^{cr}&\simeq&0.65-0.35(\gamma_0-4.2).
\end{eqnarray}

\begin{figure}
\narrowtext
\epsfxsize=3.3truein
\vbox{\hskip 0.05truein
\epsffile{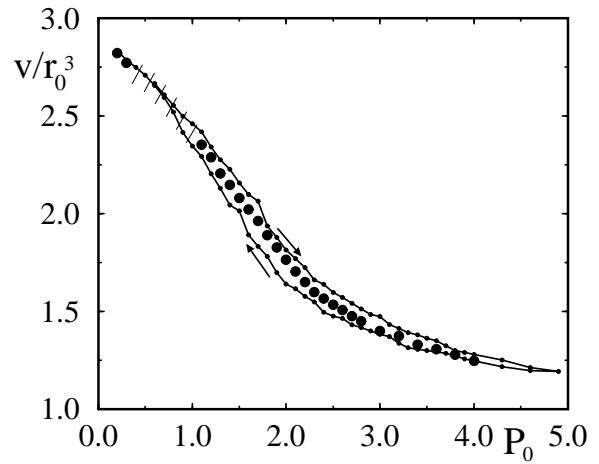}}
\medskip
\caption{Specific volume as a function of $P$ at $T=0$ from the
simulations shown in Fig. \ref{vt1} (large symbols) and from a single run 
at $T=0.01$
(20000 steps per temperature) increasing and decreasing $P$ (small 
symbols. The path of the simulation is indicated by the arrows). The
dashed region contains no large symbols since here crystallization could  
not be avoided when reducing $T$. Note that the limiting values of $v/r_0^3$
at very high and low $P$ are respectively 0.808 and 2.727.}
\label{vdept0}
\end{figure}

The locus of the anomalies of $v$ and $K_T$ also move with $\gamma$. We
note that if $\gamma$ is such that the critical point exists, the line of 
$K_T$ maxima  necessarily ends at the critical point
(since at the critical point $K_T\rightarrow\infty$). For
the extrema of $v$ this is not necessarily so, although it is known that
the anomalies in $K_T$ and $v$ are thermodynamically
related.\cite{stanley4}

\section{Characteristics of melting}

In this section we show results of numerical simulations that focus on
the melting of the most expanded of the solid phases of our
system\cite{jaglapre} (which is the equivalent of ice Ih in
water). 
Figure \ref{vt2} shows the specific volume $v$ as a function of $T$ for
different values of $P$ obtained in slow simulations
decreasing and increasing $T$ in a system of 216 particles. The hysteresis
upon heating and cooling embraces the position of the thermodynamic
melting transition
temperature. In all the range of $P$ indicated in this figure, the system
freezes into one and the same solid configuration, corresponding to a
dense
stacking of triangular planes, in which each particle has twelve
nearest neighbors at distance $r_1$ (the dispersion in the limiting value
of $v$ when $T\rightarrow 0$ is due to a few defects that remain in the
solid structure). Upon increasing $T$, $v$ increases for $P_0\lesssim
0.95$
(which is of course the standard behavior), but
decreases for larger $P_0$. This decrease is driven by the possibility of
particles of being at distances smaller than $r_1$ from each other.
Depending on $P$, the tendency of particles to become closer (gaining energy
from the $Pv$ term) may be higher than the entropic tendency to increase
$v$.  
In the same way, at the lowest pressures, the solid melts by increasing its
volume, whereas at the largest pressures shown in Fig.  
\ref{vt2} it melts by reducing its
volume. This is consistent with the form of the solid-liquid border in the
$P$-$T$ plane that  is seen in Figure \ref{pt4}, which has positive
derivative at low $P$, but has negative derivative at larger $P$. 

\begin{figure}
\narrowtext
\epsfxsize=3.3truein
\vbox{\hskip 0.05truein
\epsffile{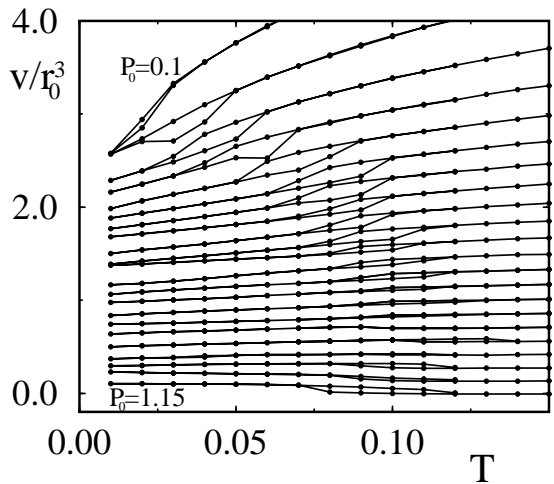}}
\medskip
\caption{The specific volume $v$ as a function of $T$ for different values
of $P_0$, from 0.1 to 1.15, in steps of 0.05 for a system of 216 particles.
Each curve was
vertically displaced by a term $-2P_0$ to allow a better visualization.  
Hysteresis is the result of successive cooling and heating. Note the 
normal melting at low $P_0$, and the anomalous one for $P_0{\protect
\gtrsim} 0.95$.}
\label{vt2}
\end{figure}

\begin{figure}
\narrowtext
\epsfxsize=3.3truein
\vbox{\hskip 0.05truein 
\epsffile{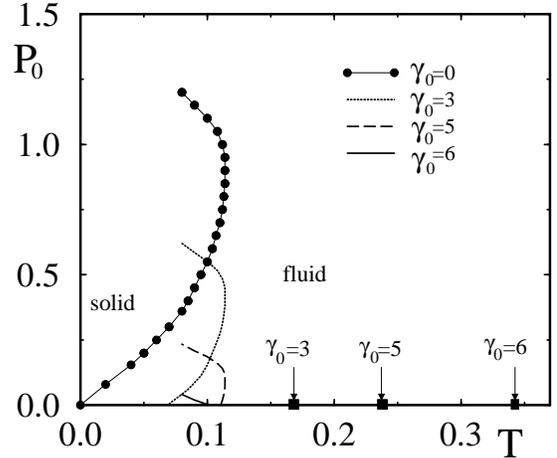}}
\medskip
\caption{The melting line of the lowest pressure solid structure, for
$\gamma_0=0$, 3, 5, and 6. There is also a liquid-gas coexistence line
for all $\gamma_0 \neq 0$ that in the scale of the figure cannot be
distinguished from the $P_0=0$
axis. This line ends in
the critical point indicated for each case by a square.} 
\label{pt4}
\end{figure}

In the
same Figure \ref{pt4} we see the modification of the phase diagram when
we consider the van der Waals attraction. Any finite value of $\gamma$ 
makes a liquid-gas first order line appear.\cite{hemmerleb}
In the scale of Fig. \ref{pt4} this critical line cannot be distinguished from
the $P=0$ axis, only the critical temperature is indicated.
In addition, the whole solid-fluid coexistence
line basically moves down with $\gamma_0$. If $\gamma_0 \lesssim 5.5$
the triple point that is defined is
`standard', in the sense that the slope of the solid-liquid coexistence line
is positive at the triple point. For larger $\gamma_0$ the triple point is
`anomalous' (the slope of the solid-liquid coexistence line
is negative).

\section{Summary and conclusions}

We studied the phase diagram of a model of spherical particles
with pairwise interactions, consistent of a hard core at a distance
$r_0$ plus a repulsive linear shoulder that extends up to distance $r_1$.
This potential favors the particles to be in one or
the other (depending on $P$) of the two different equilibrium distances $r_0$ 
and $r_1$.
On top of that, a long range van der Waals attraction was also included.

The solid phase of the system exhibits polimorfism. Namely, there are
different sectors
of the $P$-$T$ phase diagram in which the crystalline structure of the
system is different. This
behavior is observed even in the case of non attractive part in the
potential (i.e., $\gamma=0$).

The fluid phase part of the phase diagram has the following
characteristics. At low pressures particles prefer to be at
distances $\sim r_1$ from each other, whereas at high pressures the typical
distance is $r_0$ ($<r_1$). This implies a crossover region of $P$ with an
anomalously large isothermal compressibility $K_T$.
When we include the van der
Walls attraction ($\gamma\neq 0$) the anomaly in $K_T$ may become (if
$\gamma$ is sufficiently large) a first order transition line (similar to
the liquid-gas coexistence line). This first order line starts
from a finite $P$ at $T=0$, and ends in a critical point at finite $T$ and
$P$. From this point
the line of $K_T$ maxima continues towards larger values of $T$. There are
also anomalies in the density of the system, which has extrema in a locus
that, whereas it does not necessarily touch the critical point, appears in
the region that is influenced by the existence of it.

For $\gamma=0$, the melting line of the most expanded solid structure
in the $P$-$T$ plane has positive derivative at low pressures, but is
reentrant at higher $P$.
This reentrant behavior is associated (through the Clausius-Clapeyron
equation) to a melting with density increasing. When the van der Waals
attraction is included, a liquid-gas first order line appears, that ends in
a critical point as usual. This liquid-gas line defines a triple
point where it meets the fluid-solid line. For small $\gamma$ the slope of
the solid-liquid line at the triple point is positive, but it becomes
negative if $\gamma$ is large enough.

Our model, although very simple, has  many of the
properties that characterize water as an anomalous fluid,
and gives insight into the properties of real water.
Actually, the simplicity of the model allows to single out the crucial
characteristic that produces all the anomalies, without the complications
introduced by non-spherical interactions and cooperative hydrogen
bonding in real water. 
This characteristic
is the existence in the interatomic potential of two different equilibrium
distances for the particles. 
From all our results it is difficult to elude the claim that
there must be an
effective description of the interaction in water in which two diggerent
distances compete as being the most stable one. In fact, it would be 
even more daring to say that all the similarities we found are accidental. 
Although there is
evidence favoring this view,\cite{stanleyrev,cho,campolat} the complications
added by the peculiarities of water molecules has made this point very
disputed.\cite{velascocho} 

Among all anomalous properties of water, the existence of the
second critical point is the one that is not fully proven to ocurr, 
and also the one that has been most elusive to adress numerically
in previous studies. 
Our model shows that its existence is a consequence of the effect
of the attractive part of the potential on a system that (due to
peculiarities of the interaction) possesses anomalously large values of
$K_T$ at some pressure. At this point experimental evidence about the
existence of two amorphous phases (LDA and HDA) that transform reversibly
into each other seems crucial to indicate that in water, the attraction
between molecules is strong enough to bring the second critical
point into existence. 

From a more fundamental point of view,
we note that our model has essentially two free
parameters, the ratio between equilibrium distances $r_1/r_0$ and the
strength of the van der Waals attraction $\gamma$. Other characteristics,
as if the ramp between $r_0$ and $r_1$ is linear or not, are only marginal
for the phase diagram that is obtained.\cite{jaglapre} An important
result of
our study is the fact that these two parameters determine the phase
behavior of the fluid phase both in the zone where the fluid is stable,
and also in the deeply supercooled region. If water admits a similar
effective representation in term of two parameters,\cite{nota2} then these
could be extracted from fitting experimental data in the high temperature
region, and then used, relying on our model, to predict the supercooled
part of the phase
diagram, in particular the existence and location of the second
critical point. This means that the present model may even be of
quantitative importance. Work on this direction is under way.


\begin{references}
\bibitem{nota}In order to use Eq. (\ref{algo}), it is still neccesary to know
the integration constant when we determine $s_{HS}$ from (\ref{sp}) and
(\ref{algo0}). This
constant is physically related to the fact that we have divided the total
entropy in a vibrational ($s^{HS}$) and a combinatorial ($s^I$) part. Since a
precise determination of this constant is difficult, and we pursue a
qualitative description of the phase diagram, we took this constant to be
zero, having verified that finite values do not change the qualitative
characteristics of the results.
\bibitem{nota2}In fact, fitting of properties of water in the stable and
metastable regime with expressions derived from a Landau-type theory with
only one free parameter is remarkably good (see Ref. [12]).
\end{references}
\end{document}